\documentclass[structabstract]{aa}  

\usepackage{graphicx}
\usepackage{txfonts}
\usepackage{dsfont}
\usepackage{natbib,twoopt}
\usepackage{upgreek}
\usepackage{multirow}

\usepackage[breaklinks=true]{hyperref}	
\hypersetup{
    bookmarks=true,         
    unicode=false,          
    pdftoolbar=true,        
    pdfmenubar=true,        
    pdffitwindow=false,     
    pdfstartview={FitH},    
    pdftitle={The velocity field of sunspot penumbrae II. Return flow and magnetic fields of opposite polarity},    
    pdfauthor={Morten Franz},     
    pdfsubject={A&A Paper},   
    pdfcreator={Morten Franz},   
    pdfproducer={Morten Franz}, 
    pdfkeywords={Sunspots} {Penumbra} {Spectropolarimetry} {HINODE}, 
    pdfnewwindow=false,      
    colorlinks=true,       
    linkcolor=black,          
    citecolor=blue,        
    filecolor=blue,      
    urlcolor=blue,           
    linkbordercolor={1 0 0}, 	
    citebordercolor={0 1 0}, 
    urlbordercolor={1 1 1},
    linktocpage=false,
    raiselinks=true,
    hyperfootnotes=false
}

 \bibpunct{(}{)}{;}{a}{}{,} 
 \newcommandtwoopt{\citeads}[3][][]{\href{http://adsabs.harvard.edu/abs/#3}{\citealp[#1][#2]{#3}}} 
 \newcommandtwoopt{\citepads}[3][][]{\href{http://adsabs.harvard.edu/abs/#3}{\citep[#1][#2]{#3}}} 
 \newcommandtwoopt{\citetads}[3][][]{\href{http://adsabs.harvard.edu/abs/#3}{\citet[#1][#2]{#3}}}
 \newcommandtwoopt{\citeyearads}[3][][]{\href{http://adsabs.harvard.edu/abs/#3}{\citeyear[#1][#2]{#3}}}

\begin{document}

   \title{The velocity field of sunspot penumbrae}
   
   \subtitle{II. Return flow and magnetic fields of opposite polarity}

   \author{M. Franz \and R. Schlichenmaier}

   \institute{Kiepenheuer Institut f\"ur Sonnenphysik,
              Sch\"oneckstra\ss{}e 6, D-79104 Freiburg\\
              \email{morten@kis.uni-freiburg.de, schliche@kis.uni-freiburg.de}}

   \date{Submitted: Nov 8, 2012, Accepted: Dec 19, 2012}

\abstract
{}
{We search for penumbral magnetic fields of opposite polarity and for their correspondence with downflows.}
{We used spectropolarimetric HINODE data of a spot very close to disk center to suppress the horizontal velocity components as much as possible. We focus our study on 3-lobe Stokes V profiles.}
{From forward modeling and inversions, we show that 3-lobe profiles testify to the presence of opposite magnetic fields. They occur predominately in the mid and outer penumbra and are associated with downflows in the deep layers of the photosphere.}
{Standard magnetograms show that only 4\% of the penumbral area harbors magnetic fields of opposite polarity. If 3-lobe profiles are included in the analysis, this number increases to 17\%.}


   \maketitle
   

\section{Introduction}

Our knowledge of the penumbral magnetic and velocity field has increased tremendously \citep[see e.g. ][]{2011LRSP....8....3R,2011LRSP....8....4B}. The discrepancy between the inclination of the average magnetic field and the observed velocity field is one example. While a zenith angle of less than 90$^{\circ}$ is deduced for the azimuthally averaged magnetic field \citep{1993ApJ...418..928L}, the Evershed flow \citep{1909MNRAS..69..454E} appears parallel to the surface throughout the penumbra \citep{1964ApNr....8..205M}. High resolution observation show that the inclination of the penumbral magnetic field is larger within the dark penumbral filaments, e.g. \citet{1969SoPh...10..384B,1992A&A...264L..27S,1993ApJ...403..780T}. Today it is accepted that the inclination of the penumbral magnetic field has strong fluctuations on scales of less that 1$\arcsec$ \citep{1993A&A...275..283S,2000A&A...361..734M,2004A&A...427..319B}. It was further argued that the Evershed flow occurs along the dark penumbral structures, where the field is more parallel to the solar surface and possibly weaker \citep{1997ApJ...477..485S,2000A&A...364..829S,2005A&A...436.1087L}. 

However, the correct mode of penumbral magnetoconvection is still a controversy. Evidence exists that the upflows in the inner penumbra are the source of this flow, while the downflows in the outer penumbra are the regions where it returns below the solar surface \citep{2000A&A...358.1122S,2007PASJ...59S.593I,2009A&A...508.1453F}. This is in accordance with the `flux-tube' model \citep{1968MitAG..25..194M,1997Natur.390..485M,1998A&A...337..897S} where penumbral plasma is channeled along bundles of strong magnetic field. This model, in turn, requires a large amount of magnetic field submerging with these downflows. This should show up as a significant amount of opposite polarity signal in penumbral magnetograms. There have been reports of magnetic fields returning within the penumbra \citep{1997Natur.389...47W,2001ApJ...547.1130W,2001ApJ...549L.139D}, but even in magnetograms of high spatial resolution; e.g.,~\citep{2005A&A...436.1087L} the amount of return flux is far less than what is expected from penumbral `flux tube' models \citep{2003A&ARv..11..153S,2003A&A...411..257S}. In the debate about the correct mode of magneto convection and the penumbral structure, this has been taken as an argument to support the `gappy' penumbral model \citep{2006A&A...460..605S,2006A&A...447..343S}. However, \citet{2007PASJ...59S.593I} used HINODE data to compute magnetograms in the far line wing and found several opposite polarity patches that are associated with penumbral downflows. Nevertheless, detailed studies of the amount of penumbral return flux are still missing.

In this contribution we study 3-lobe Stokes V profiles, which indicate the presence of magnetic fields of opposite polarity, if the spot is observed at disk center. Considering these profiles, we find more opposite polarity than previously reported. We use spots that are as close as possible to disk center, and compute maps of various physical quantities in Sec.~\ref{sec:obs}. We discuss 3-lobe V-profiles that have an extra redshifted lobe in the red wing and V-profiles that have a blue-shifted hump in the blue wing. We use geometrical models (Sec.~\ref{sec:tlobprof}) and inversion techniques (Sec.~\ref{sec:inversion}) to model these profiles. We conclude that blue-hump profiles are location of upflows with the spot polarity, and 3-lobe profiles are locations of opposite polarity. We find that 3-lobe profiles preferentially occur at locations of downflows (Sec.~\ref{sec:shape}). In Sec.~\ref{sec:distrib} we demonstrate that `classical' magnetograms detect only a small fraction of opposite polarity, since they do not consider 3-lobe profiles. We quantify how the 3-lobe profiles are distributed throughout the penumbra. In Sec.~\ref{sec:discuss} we discuss our results. 
Summary and conclusion are given in Sec.~\ref{sec:conclusions}.

 
\section{Observation \& data processing}
\label{sec:obs}

We used data obtained by the spectropolarimeter (SP) \citep{2001ASPC..236...33L} of the solar optical telescope (SOT) \citep{2008SoPh..249..167T} onboard the {\it{Hinode}} satellite. The SP records the Stokes spectra of the two iron lines at 630.15 nm and 630.25 nm, with Land\'e factors of $\rm{g}_{\rm eff}=1.67$ and $\rm{g}_{\rm eff}=2.5$, respectively. By scanning the spectrograph slit across the target in steps of 0."15 and with a wavelength sampling of 2.15 pm/pixel, two-dimensional maps of the field of view (FOV) were obtained. Since the width of the slit is equivalent to 0."16 and the pixel size along the slit is 0."16 on average, normal SP scans provide a spatial resolution of 0."32 \citep{2009ASPC..415..323C}. For an exposure time of 4.8 s per slit position, the 1$\upsigma$ noise level in the Stokes spectra is around $10^{-3}{\cdot}\rm{I}_{\rm{c}}$, where ${\rm{I}}_{\rm{c}}$ is the continuum intensity. 

\begin{table}[h]
\begin{center}
	\caption{Data samples used in this study.}
	\begin{tabular}{cccc}
		\hline
		\hline
		\\[-2ex]
		{Name}&{NOAA} & {Date of} & {Heliocentric}\\
		{}&{Active region} & {observation} & {angle $\Uptheta$ [$^{\circ}]$}\\
		\hline
		\\[-2ex]
		Spot A & {10923} & {Nov 14$^{\rm{}th}$ 2006} & {10.8 - 5.4}\\
		Spot B & {10923} & {Nov 14$^{\rm{}th}$ 2006} & {12.1 - 3.9}\\
		Spot C & {10930} & {Dec 11$^{\rm{}th}$ 2006} & { 9.8 - 3.6}\\
		Spot D & {10933} & {Jan 05$^{\rm{}th}$ 2007} &{4.8 - 1.5}\\
		\hline
	\end{tabular}
	\label{Tab_1}
\end{center}
\end{table}

For our study, we used the raw data summarized in Table~\ref{Tab_1} and reduced it, using the IDL routine `sp\_prep.pro' provided with the `Solar Soft' package. In Sec. \ref{sec:shape} and \ref{sec:distrib} we focus on dataset Spot D, because it is the one located closest to disk center. From the Stokes spectra, we computed maps of:
\begin{itemize}
\item{continuum intensity: $\rm{I}=\rm{I}_{\rm{c}}/\rm{I}_{\rm{qs}}$ where $\rm{I}_{\rm{qs}}$ is the average continuum intensity of the quiet Sun}
\item{linear polarization: $\rm{Q}(\uplambda)=\hat{\rm{Q}}(\uplambda)/\rm{I}_{\rm{c}}$ and $\rm{U}(\uplambda)=\hat{\rm{U}}(\uplambda)/\rm{I}_{\rm{c}}$ with $\hat{\rm{Q}}(\uplambda)$ and $\hat{\rm{U}}(\uplambda)$ as the measured Stokes Q and U profiles}
\item{circular polarization: $\rm{V}(\uplambda)=\hat{\rm{V}}(\uplambda)/\rm{I}_{\rm{c}}$} where $\hat{\rm{V}}(\uplambda)$ represents the measured Stokes V profile
\item{total polarization: $\rm{P}_{\rm{tot}}=\int [(\rm{Q(\uplambda)}^{2}+\rm{U(\uplambda)}^{2}+\rm{V(\uplambda)}^{2})]^{1/2}\rm{d\uplambda}$}
\item{net circular polarization\footnote{For technical reasons the integral is taken over the entire spectral region of the HINODE SP covering both Fe lines. This quantity has also been termed broad band polarization \citep[e.g.][]{1975A&A....41..183I}.}: $\mathcal{N} = \int{\rm{V(\uplambda)d}\uplambda}$}
\item{area asymmetry : $\mathcal{A} = \mathcal{N}/\!\!\int|\rm{V}(\uplambda)|\rm{d}\uplambda$}
\item{Doppler velocity along the line of sight: ${\rm{v}}_{\rm{dop}}$}
\item{ and a mask of 3-lobe Stokes V profiles: V$_{\rm3-lobe}$} 
\end{itemize}

The calculation of the maps of I, V, $\rm{P}_{\rm{tot}}$, $\mathcal{N}$ and $\mathcal{A}$ is straightforward. The map of ${\rm{v}}_{\rm{dop}}$ is constructed from the shift of the line wing with respect to a reference wavelength. We refer the reader to \citet{2009A&A...508.1453F} for a detailed explanation on the calculation of ${\rm{v}}_{\rm{dop}}$ including an absolute calibration of the wavelength scale. 

The mask of 3-lobe Stokes V profiles (V$_{\rm3-lobe}$) is constructed from the Fe 630.25~nm line, because it is a simple line triplet and has the higher magnetic sensitivity. This makes it easier to identify the third lobe on the red side of the regular Stokes V profile. As indicated by the colored regions in the upper right hand panel of Fig.~\ref{fig:Franz_fig00}, we split the spectrum between 630.205~nm and 630.325~nm into a sequence of three windows. The first two windows (blue and orange) have a bandwidth of 20~pm, while the third one (red) is 45~pm wide. This sequence of spectral windows is scanned as a whole across the Fe I 630.25~nm line.  To identify 3-lobe Stokes V profiles, a number of conditions have to be fulfilled. a) The lobes in the blue (left) and red (right) windows are positive. b) The maxima of the lobes and the spectral points next to it are positive and above the $3\sigma$ noise level. c) The lobe of the V profile in the orange (middle) window is negative. d) The minimum itself and the spectral measurements next to it are negative and above the $3\sigma$ noise level.
The requirement that the neighboring spectral measurements have to be above the $3\sigma$ threshold and have the same sign as the extrema was applied to exclude artifacts caused by the magneto-optical effect, i.e. additional peaks around the line core position in Stokes V. A similar method for automatic detection of blue-hump profiles encounters difficulties and has not been devised.


\section{Blue-hump and 3-lobe Stokes V profiles}
\label{sec:tlobprof}

There has been a long history of studies of the asymmetry in Stokes V profiles. \citet{1951ApJ...114....1B} already noted differences in intensity and width between the two components of circular polarized light from a variable star. He termed this phenomenon the crossover effect\footnote{We use the term 3-lobe profiles instead of crossover effect to differentiate these profiles from stellar profiles.}. \citet{1972SoPh...22..119G} were among the first to observe Stokes V profiles with more than two lobes in the spectra of sunspots. By systematically analyzing the asymmetry of Stokes profiles \citet{1992ApJ...398..359S} and \citet{2002A&A...381..668S} found that Stokes V profiles with more than two lobes predominate, but are not exclusive \citep{2002NCimC..25..543B}, along the magnetic neutral line.

\begin{figure}[h!]
	\centering
		\includegraphics[width=\columnwidth]{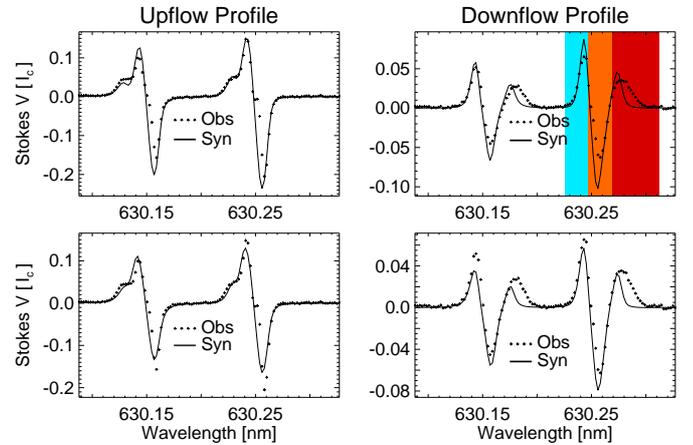}
		\caption{Stokes V profiles from typical penumbral up- and downflows observed with HINODE (crosses). Top row: Synthetic profiles (solid) radiative transfer calculations within a {\it two-layer atmosphere}. The colored regions in the top right plot indicate the spectral regions that are used to identify 3-lobe profiles. Bottom row: Synthetic spectra (solid) originating in a {\it two-component atmosphere}.}
		\label{fig:Franz_fig00}
\end{figure}

With HINODE these peculiar profiles are regularly measured throughout the entire penumbra. \citet{2007PASJ...59S.593I} and \citet{2010AN....331..570F} found that Stokes V profiles from penumbral upflow regions contain an additional hump on the blue side of an antisymmetric Stokes V profile (blue-hump profiles), while the corresponding profiles from downflow regions contain a third lobe of opposite sign on the red side (3-lobe profiles) -- cf.~Fig.~\ref{fig:Franz_fig00}.

There are two simple atmospheric configurations that yield blue-hump or 3-lobe Stokes V profiles. The first is a {\it two-layer atmosphere}. It includes strong jumps (or gradients) of the atmospheric parameters along the line of sight (LOS) and within the line-forming region, e.g.~\citet{1978A&A....64...67A} or \citet{1983SoPh...87..221L}. The second configuration is a {\it two-component atmosphere}. It mixes two laterally displaced but spatially unresolved atmospheres (in these atmospheres no gradient of Doppler velocity, strength, inclination, and azimuth of the magnetic field are present along the LOS) within one resolution element, e.g.~\citet{2004A&A...427..319B}.

\subsection{Geometrical models}
\label{sec:fwdmodel}

We studied the ability of these two configurations to reproduce blue-hump or 3-lobe Stokes V profiles by means of radiative transfer calculations. We used the synthesis module of SIR \citep[Stokes Inversion based on Response functions][]{1992ApJ...398..375R}, together with a modified version of the Harvard Smithsonian Reference Atmosphere (HSRA) of \citet{1971SoPh...18..347G}. The changes we introduced in the HSRA to investigate the {\it two-layer atmosphere} are listed in Table~\ref{Tab_2}.

\begin{table}[h]
\begin{center}
	\caption{{\it Two-layer atmosphere}: Modification to HSRA.}
	\begin{tabular}{cccc}
		\hline
		\hline
		\\[-2ex]
		{Optical depth}&{Field strength} & {Zenith angle} & {Velocity}\\
		{log($\uptau_{500}$)}&{B [G]} & {$\upgamma$ [$^{\circ}$]} & {v$_{\rm{dop}}$ [km s$^{-1}$]}\\
		\\[-2ex]
		\hline
		\multicolumn{4}{c}{Upflow case}\\
		\hline
		\\[-2ex]
		 0.0 to $-$0.5 & {1000} & {60} & {$-$6.5}\\
		$-$0.6 to $-$3.0 & {1500} & {20} & {0.0}\\
		\hline
		\multicolumn{4}{c}{Downflow case}\\
      		\hline
		\\[-2ex]
		 0.0 to $-$0.5 & {1000} & {120} & {8.5}\\
		$-$0.6 to $-$3.0 & {700} & {60} & {0.0}\\
		\hline
	\end{tabular}
	\label{Tab_2}
\end{center}
\end{table}

To mimic a {\it two-component atmosphere}, we performed radiative transfer calculations within 
a modified version (cf. Table~\ref{Tab_3}.) of the HSRA. To obtain the final Stokes V profile, we mixed the two Stokes V profiles from the individual atmospheric components with different filling factors.

\begin{table}[h]
\begin{center}
	\caption{{\it Two-component atmosphere}: Modification to HSRA.}
	\begin{tabular}{cccc}
		\hline
		\hline
		\\[-2ex]
		{Filling factor of}&{Field strength} & {Zenith angle} & {Velocity}\\
		{component [\%]}&{B [G]} & {$\upgamma$ [$^{\circ}$]} & {v$_{\rm{dop}}$ [km s$^{-1}$]}\\
		\\[-2ex]
		\hline
		\multicolumn{4}{c}{Upflow case}\\
         \hline
         	\\[-2ex]
		30 & {1000} & {60} & {$-$6.5}\\
		70 & {1500} & {20} & {0.0}\\
		\hline
		\multicolumn{4}{c}{Downflow case}\\
		\\[-2ex]
         \hline
		30 & {1000} & {120} & {8.5}\\
		70 & {700} & {60} & {0.0}\\
		\hline
	\end{tabular}
	\label{Tab_3}
\end{center}
\end{table}

Figure~\ref{fig:Franz_fig00} summarizes the outcome of our experiment. It can be seen that both the {\it two-layer atmosphere} and the {\it two-component atmosphere} is able to reproduce blue-hump and 3-lobe Stokes V profiles. Keep in mind that these models are constructed to reproduce distinct features and not to minimize the difference between the observed and synthetic profiles. The latter is done in Sec.~\ref{sec:inversion} by applying an inversion method.

We found a vast range of possible atmospheric configurations that yield blue-hump or 3-lobe Stokes V profiles. However, in all these realizations it is necessary to introduce a) magnetic fields of sufficient strength in both components or layers, b) an unshifted and a blueshifted component or layer for the upflow case, c) an unshifted and a redshifted component or layer for the downflow case, d) magnetic fields of the same polarity in the blueshifted component or layer, and e) magnetic fields of {\it opposite polarity} in the redshifted component or layer.

In both configurations magnetic fields of opposite polarity are present in downflow regions. In the {\it two-component atmosphere} they simply lie next to each other in one resolution element. In the {\it two-layer atmosphere} the downflows are concentrated in the lower layers of the photosphere, $0 \le \rm{log}(\uptau) \le -0.5$, where the magnetic field returns below the solar surface,  $\upgamma > 90^{\circ}$. In the layers above, $0.6 \le \rm{log}(\uptau) \le -3.0$,  the atmosphere is at rest and the magnetic field has the same polarity as the umbra.

\subsection{Asymmetry of blue-hump and 3-lobe profiles}
\label{sec:asym}

So far our results have not allowed determining whether gradients along the LOS or laterally unresolved structures within a resolution element are at the origin of blue-hump and 3-lobe profiles. However, it is possible to use the net circular polarization ($\mathcal{N}$) or the area asymmetry\footnote{In contrast to $\mathcal{N}$, $\mathcal{A}$ is normalized to the area of the profile. This makes the latter more appropriate when comparing the asymmetry of Stokes V profiles with different amplitudes.} ($\mathcal{A}$) to distinguish between these two possibilities. According to \citet{1978A&A....64...67A}, a gradient of v$_{\rm {dop}}$ along the LOS is not only necessary, but sufficient to produce non-vanishing values of $\mathcal{N}$, hence $\mathcal{A}$. If a gradient in v$_{\rm {dop}}$ is present, gradients of other parameters, i.e. the strength, the inclination or the azimuth of the magnetic field, alter the amount of $\mathcal{A}$ \citep{1996SoPh..164..191L}.

For the two profiles presented before we measure $\mathcal{A}^{\rm{obs}}_{{\rm{up}}}=19\%$ and $\mathcal{A}^{\rm{obs}}_{{\rm{down}}}=84\%$, respectively. Such high values of $\mathcal{A}$ are common for Stokes V profiles from penumbral up- or  downflows and need to be reproduced by the synthetic profiles.

The {\it two-layer atmosphere} produces a significant amount of area asymmetry: $\mathcal{A}^{\rm{layer}}_{{\rm{up}}}=-24\%$ and $\mathcal{A}^{\rm{layer}}_{{\rm{down}}}=24\%$. However, the synthetic upflow profile yields the wrong sign of $\mathcal{A}$ when compared to observations. 

The {\it two-component atmosphere} is unable to produce any area asymmetry, since the atmospheric parameters do not change along the LOS. Therefore, we do not consider this configuration in the following. 

 
 \section{Inversion}
\label{sec:inversion}
 
We used the inversion tool of the SIR code \citep{1992ApJ...398..375R} to create an atmosphere as simple as possible, but still capable of reproduceing all spectral features (blue-hump and 3-lobe), as well as the area asymmetry of the measured Stokes V profiles. In contrast to forward modeling, this technique minimizes the differences between observed and synthetic Stokes profiles, yielding a maximal goodness of the fit. In SIR, physical parameters are interpolated between `nodes', which are placed in the atmosphere. In the iteration process, the values at the nodes are changed. If there is only one node for a parameter, its initial function along the line-of-sight is shifted according to the change in the node. When the iteration process converges, a subsequent iteration loop (i.e. cycle), with an increased number of nodes, can be processed.
 
\subsection{Setup and procedure}
\label{sec:setup}

\begin{figure}[t!]
	\centering
		\includegraphics[width=\columnwidth]{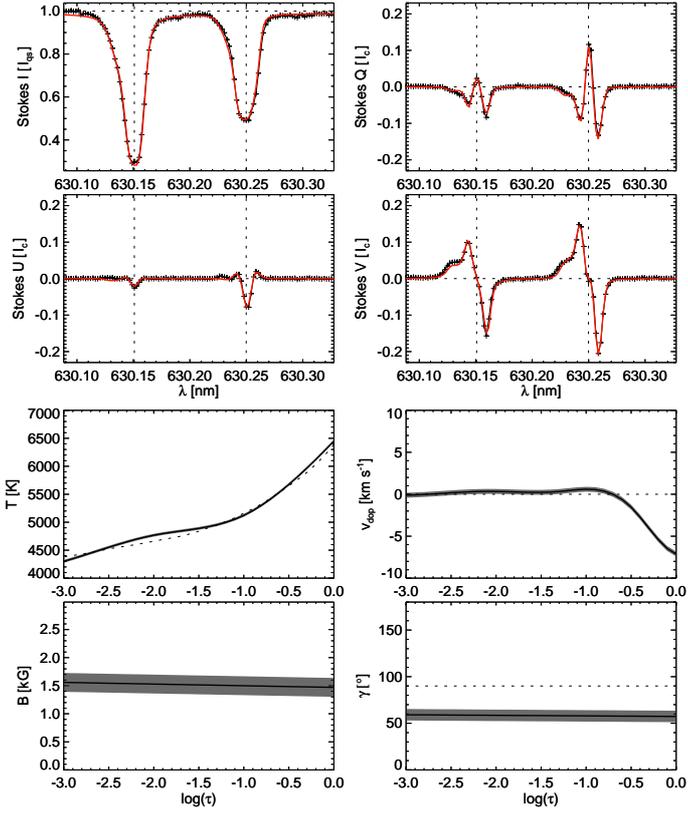}
		\caption{Top part clockwise: Representative Stokes I,  Q,  V, and U profiles (black) from an upflow region. The red lines indicate the result of radiative transfer calculations within the atmosphere plotted below and the dashed vertical lines represent the respective vacuum wavelength of the transition. Bottom part clockwise: Temperature, Doppler velocity, zenith angle, and magnetic field strength (black) with errors (shaded gray) at different optical depth values. The dashed lines indicate the temperature stratification in the HSRA model, plasma at rest, and a zenith angle of 90$^{\circ}$ in the respective panels.}
		\label{fig:Franz_fig01}
\end{figure} 
 
We performed a two-cycle inversion and used a modified version of the HSRA as an initial model. We randomized the temperature (T) distribution around the values of that of the HSRA, while the values for the electron pressure (p$_{\rm{e}^-}$) and microturbolence (v$_{\rm{mic}}$) were kept. We added random values of Doppler velocity (v$_{\rm{dop}}$), magnetic field strength (B), zenith angle ($\upgamma$), and azimuth ($\upphi$), but ones that are constant in optical depth ($\uptau$). To ensure that the final atmosphere does not represent a local minimum of the $\upchi^2$ manifold\footnote{$\upchi^2$ is a weighted difference between all observed and synthetic Stokes parameters divided by the number of free parameters.}, the inversion of a given Stokes vector was performed 100 times with randomized input model atmospheres. From the resulting pool of solutions, the fit with a minimal $\upchi^2$ was selected.
 
In the first cycle, we used one node in all atmospheric parameters except for p$_{\rm{e}^-}$, which remains the same as in HSRA. The macroturbolence (v$_{\rm{mac}}$) is also a free parameter, but we did not consider a straylight component. This leads to seven degrees of freedom (DOF). 
 
In the second cycle, we allowed for three nodes in the stratification of T and two in that of B and $\upgamma$, while the rest was not changed. For v$_{\rm{dop}}$, it was necessary to allow for seven nodes to be able to retrieve the shape of the observed Stokes V profiles. In fact, it is impossible to model the additional hump in upflow profiles with less DOF in v$_{\rm{dop}}$. This is feasible for the red-shifted profiles, but only at the cost of increasing the number of nodes in B and $\upgamma$ to three. The reason for this lies in the mode of operation of the code itself. SIR distributes the nodes equally with $\uptau$, which makes it difficult to model a flow mainly present in the deep atmospheric layers. All together there are 17 DOF in the second cycle.

\subsection{Upflow atmosphere}
\label{sec:upflow case}

\begin{figure}[t!]
	\centering
		\includegraphics[width=\columnwidth]{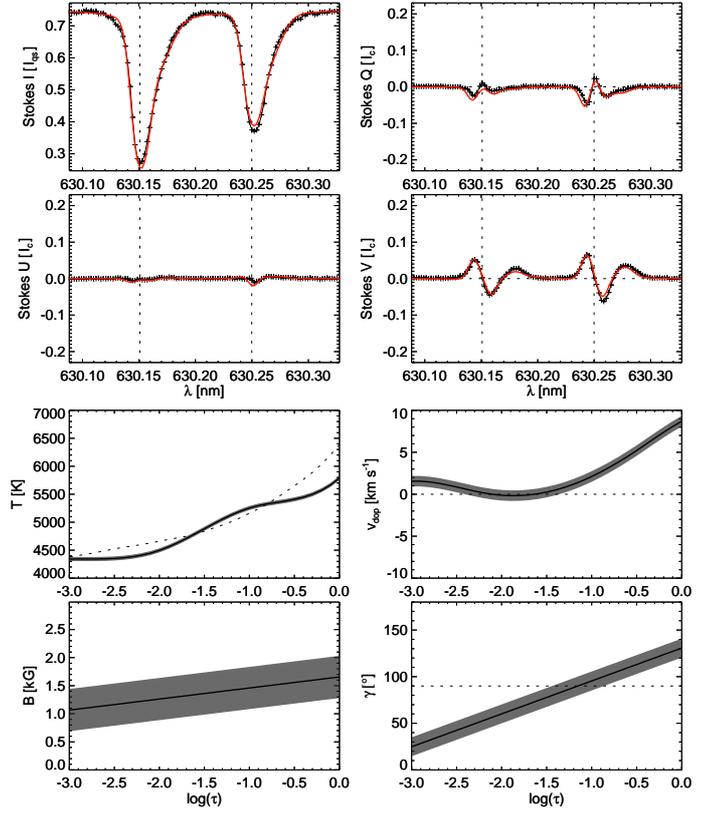}
		\caption{Same as Fig.~\ref{fig:Franz_fig01} but for a representative profiles from a downflow region. The magnetic field changes its polarity in the deep photospheric layers.}
		\label{fig:Franz_fig02}
\end{figure}

Figure~\ref{fig:Franz_fig01} represents the solution for one upflow case. The four panels at the bottom illustrate the stratification of the atmospheric parameters. Depicted are T, v$_{\rm{dop}}$, B, and $\gamma_{\rm{mag}}$ only, because the other parameters where either not inverted (p$_{\rm{e}^-}$) or remain constant throughout the line-forming region (v$_{\rm{mic}}$, v$_{\rm{mac}}$, and $\upphi$). The variation in the atmospheric parameters along the LOS is plotted between log($\uptau)=-3$ and log($\uptau)=0$, where the sensitivity of the two lines is highest.

In this model, T increases monotonously from 4300~K at log($\uptau)=-3$ to 6450~K at log($\uptau)=0$, with an error of $\pm$30~K. Except for a little plateau around {log$(\uptau)=-2$}, this distribution is similar to that of the HSRA (cf.~dashed line in the respective plot of Fig.~\ref{fig:Franz_fig01}). The strongest flows occur in the low layers of the atmosphere, where the largest amplitudes of v$_{\rm{dop}}$ are present. Between log$(\uptau)=-0.7$ and the surface, the amplitude increases from v$_{\rm{dop}} = 0$~km~s$^{-1}$ to v$_{\rm{dop}} = -7$~km~s$^{-1}$ while, within the uncertainties, almost no v$_{\rm{dop}}$ is measured in the layers above ($-3.0 \le \rm{log}(\uptau) \le -0.7$). B and $\gamma_{\rm{mag}}$ remain almost constant throughout the atmosphere with values of 60$^{\circ}\pm5.5^{\circ}$ and $1.5\pm0.16$~kG respectively.

\subsection{Downflow atmosphere}
\label{sec:downflow case}

\begin{figure*}[t!]
\begin{center}
    \includegraphics[width=\textwidth]{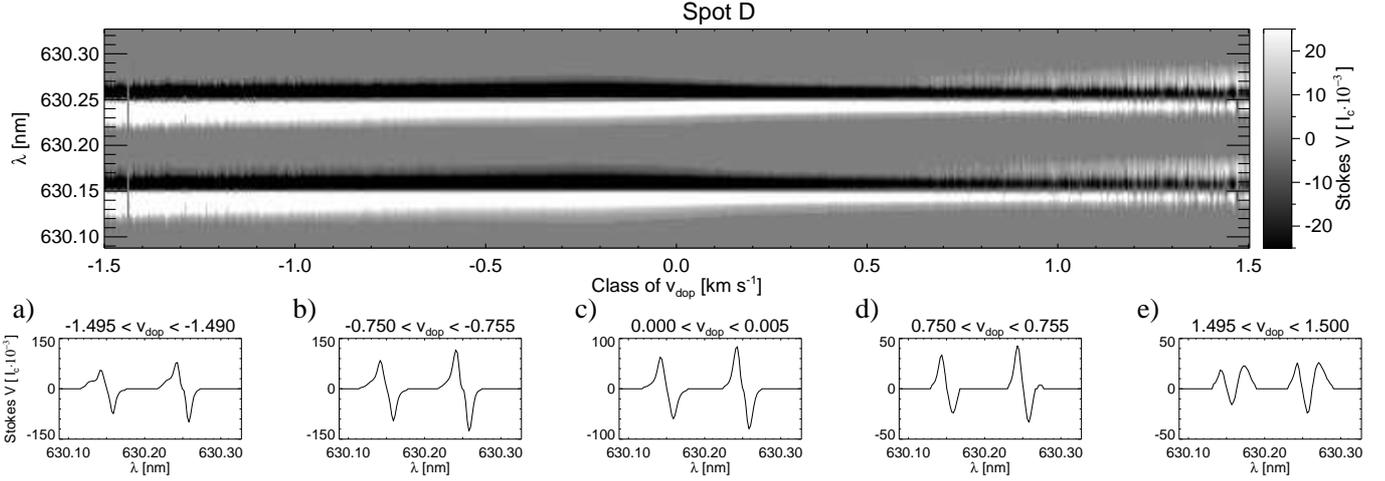}
\caption{Top: Average Stokes V profiles of different bins of Doppler velocity for the penumbra of dataset Spot D. White represents the positive lobe, black the negative lobe. The picture is saturated at a continuum intensity of 2.5\% and all measurements below the 3$\cdot\upsigma$ noise level were artificially set to zero. Bottom: Examples of averaged profiles in various velocity bins.}
\label{fig:Franz_fig03}
\end{center}
\end{figure*}

The solution for one downflow case is depicted in Fig.~\ref{fig:Franz_fig02}. When comparing these profiles to the upflow case (cf.~Fig.~\ref{fig:Franz_fig01}), it becomes apparent that Stokes I is shifted towards longer wavelengths, especially in the wing of the line. The value of I$_{\rm{c}}$ is below I$_{\rm{qs}}$, and the maximal amplitudes of Stokes Q, U, and V are lower than in the upflow profile.

In the resulting model, T increases monotonously from 4350~K at log($\uptau)=-3$ to 5300~K at log($\uptau)=0$, with an error of roughly $\pm$40 K. At lower layers, this atmosphere is cooler than the HSRA (T$_{{\rm{log}(\uptau)}}(-3) = 4400$~K and T$_{{\rm{log}(\uptau)}}(0) = 6400$~K), which explains the 25\% drop in I$_{\rm{c}}$. The distribution of T shows a little plateau around {log($\uptau)=-1$}, a feature not present in the HSRA. For v$_{\rm{dop}}$, strong flows are found in the low layers of the atmosphere. Their amplitude increases monotonously from v$_{\rm{dop}} = 0$~km~s$^{-1}$ at log$(\uptau)=-1.5$ to high values of v$_{\rm{dop}} = 8.5$~km~s$^{-1}$ at the surface. There is also a slight downflow in the higher layers (v$_{\rm{dop}} = 1.5$~km~s$^{-1}$ for $-3 \le \rm{log}(\uptau) \le -2.5$), but no flow for $-2.5 \le \rm{log}(\uptau) \le -1.5$. The uncertainty in v$_{\rm{dop}}$ is about $\pm0.5$~km~s$^{-1}$. B increases from 1.1~kG to 1.6~kG between log($\uptau)=-3$ and log($\uptau)=0$. The relatively large error of $\pm$0.35 kG makes it impossible to rule out a constant B throughout the atmosphere. In contrast to the upflow example, $\upgamma$ increases from 25$^{\circ}$ at log($\uptau)=-3$ to 130$^{\circ}$ at log($\uptau)=0$, with an error of $\pm$10$^{\circ}$. This means that the magnetic field reverses its polarity somewhere between log($\uptau)=-1.5$ and log($\uptau)=-0.9$, an atmospheric height reached by the strong downflows of the lower layers.

\subsection{Discussion}
\label{sec:diss-inv}

The area asymmetry of the resulting blue-hump profiles from an upflow region has an area asymmetry of $\mathcal{A}^{\rm{inv}}_{{\rm{up}}}=7\%$ (compared to $\mathcal{A}^{\rm{obs}}_{{\rm{up}}}=19\%$) and $\rm{Q_{\rm up}} = 0.89$. For the 3-lobe profile, we obtain $\mathcal{A}^{\rm{inv}}_{{\rm{down}}}=89\%$ (compared to $\mathcal{A}^{\rm{obs}}_{{\rm{down}}}=84\%$), while $\rm{Q_{\rm down}} = 0.78$. The inverted profiles that reproduce the observed features (blue-hump and 3-lobe). Differences exist in the area asymmetry, especially for the blue-hump profiles. This is because the SIR code minimize the difference between the observed and the synthetic Stokes profiles, and not between observed and synthetic $\mathcal{A}$.


\section{From blue-hump to 3-lobe Stokes V profiles}
\label{sec:shape}

So far we have focused on one example profile for the up- and downflow cases, respectively. In the following we study all Stokes V profiles, and we demonstrate that the occurrence of 3-lobe profiles increases continuously with Doppler velocity (v$_{\rm{dop}}$). To this end we focus on dataset Spot D, because it is the one located closest to disk center and penumbral up- and downflows are most visible.

\subsection{Behavior of average profiles of velocity bins}
\label{sec:behavior}

For this investigation we ordered the penumbral Stokes V profiles according to their Doppler velocity. Average profiles were computed for velocity bins of 5~m~s$^{-1}$, e.g. 750~m~s$^{-1} \le \rm{v}_{\rm{dop}} < 755$ ~m~s$^{-1}$. The result is depicted in Fig.~\ref{fig:Franz_fig03}, where the top panel shows the averaged profiles in all velocity bins between $\pm$1.5~km~s$^{-1}$. The five plots -- a) through e) -- in the bottom row represent examples of different velocity bins. We set all measurements below the 3$\cdot\upsigma$ noise level to zero, and omitted all velocity bins in which less than five Stokes V profiles were averaged.

Figure~\ref{fig:Franz_fig03} shows that the white area (the positive lobe) is wider than the black area (the negative lobe) in velocity bins representing strong upflows. This is due to the additional hump on the blue side of the regular Stokes V profile (cf.~Plot a) in Fig.~\ref{fig:Franz_fig03}. With decreasing upflow velocity, the width of the positive lobe decreases continuously as the additional hump disappears (Fig.~\ref{fig:Franz_fig03}~b). At zero Doppler shift, both lobes have the same width, and the profile appears antisymmetric (Fig.~\ref{fig:Franz_fig03}~c). For velocity bins representing downflows greater than 0.6~km~s$^{-1}$ we detect an additional lobe above the 3$\cdot\upsigma$ noise level in the averaged profile around $+$25~pm from the linecore of the Fe~I~630.25~nm line. With increasing downflow velocity, this additional lobe grows in width and amplitude (Fig.~\ref{fig:Franz_fig03}~e).

Since the penumbra in Spot A, B, and C is located farther away from disk center, the projection of the horizontal Evershed outflows onto the line of sight yields Doppler shifts as well. Therefore, we conducted our analysis only in an area defined by a cone angle of $\pm30^{\circ}$ perpendicular to the line of symmetry -- i.e. a line connecting the center of the spot with the center of the solar disk. In these regions the contribution of the radial outflows to the velocity signal is always less than 50\% when compared to regions along the line of symmetry. If we correct further for the different spot polarities in the respective datasets, Spot A, B, and C (not shown here) yield the same pattern of averaged Stokes V profiles as presented in the top panel of Fig.~\ref{fig:Franz_fig03}.

\subsection{Fraction of 3-lobe profiles in velocity bins}
\label{sec:fraction}

\begin{figure}[t!]
\begin{center}
    \includegraphics[width=\columnwidth]{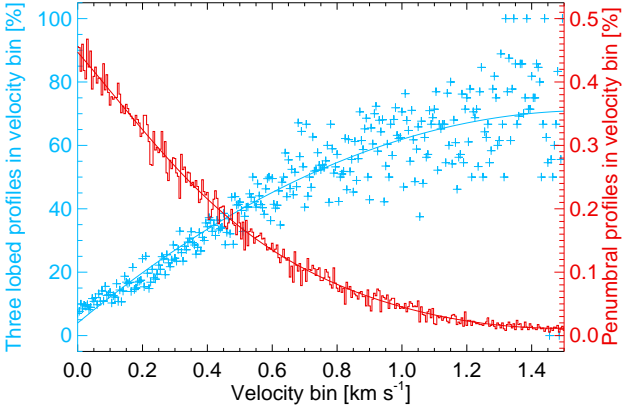}
\caption{Blue crosses and left axis: Percentage of 3-lobe profiles in bins of increasing downflow velocity for Spot D. Red histogram and right axis: Fraction of profiles in velocity bin compared to the overall number of penumbral profiles. Solid blue and solid red lines: A second order polynomial and an exponential decay fit, respectively.}
\label{fig:Franz_fig04}
\end{center}
\end{figure}

Finally we want to exclude the possibility that the additional lobe in the profiles -- panels a) through e) in Fig.~\ref{fig:Franz_fig03} -- is an artifact of the averaging process. We show that the additional lobe in the depicted profiles is not caused by a superposition of regular Stokes V profiles with various line shifts and polarities, but are due to an actual increase in the number of 3-lobe profiles in the respective velocity bin.

We use the profiles of the downflow regions, because it is easier to detect the additional lobe above the 3$\cdot\upsigma$ noise level on the red side of the profile. In each velocity bin we counted the number of profiles that obey the criteria for 3-lobe profiles -- see~Sec.~\ref{sec:obs} -- and computed their percentage on the total number of profiles. We omitted all bins with fewer than five profiles.

Figure~\ref{fig:Franz_fig04} demonstrates that the number of profiles in the respective bin decreases exponentially with velocity \citep[cf.~Fig.~3 in][]{2009A&A...508.1453F}, while the percentage of 3-lobe profiles on the overall number of Stokes V profiles in the respective velocity bin increases.

For $0.0$~km~s$^{-1} < \rm{v}_{\rm{dop}} < 0.2$~km~s$^{-1}$ 10\% to 20\% of all profiles in a velocity are 3-lobe profiles. This percentage increases almost linearly until $\rm{v}_{\rm{dop}} = 0.5$~km~s$^{-1}$  and already for $\rm{v}_{\rm{dop}} = 0.55$~km~s$^{-1}$, individual velocity bins exist in which more than 50\% of all profiles contain a third lobe. For higher downflow velocity, the scatter increases significantly, primarily because the number of profiles in a single bin decreases. According to the polynomial fit, 3-lobe profiles dominate for $\rm{v}_{\rm{dop}} > 0.7$~km~s$^{-1}$ and saturate around 70\% for $\rm{v}_{\rm{dop}} > 1.0$~km~s$^{-1}$.

Thus, it is indeed the number of 3-lobe profiles in one velocity bin that increases with downflow velocity and causes the additional lobe in the average profiles of Fig.~\ref{fig:Franz_fig03}. Note that 3-lobe profiles appear in the bins of small downflow velocity (0.0~km~s$^{-1} \le \rm{v}_{\rm{dop}} \le 0.5$~km~s$^{-1}$) and sometimes even in bins of upflows. This could be due to the spatial resolution of the instrument, which can result in a mixing of blue-hump and 3-lobe profiles from nearby up- and downflows, cf. Sec.~\ref{sec:discuss}.



\section{Distribution of penumbral 3-lobe profiles}
\label{sec:distrib}

\begin{figure*}[t]
\begin{center}
    \includegraphics[width=\textwidth]{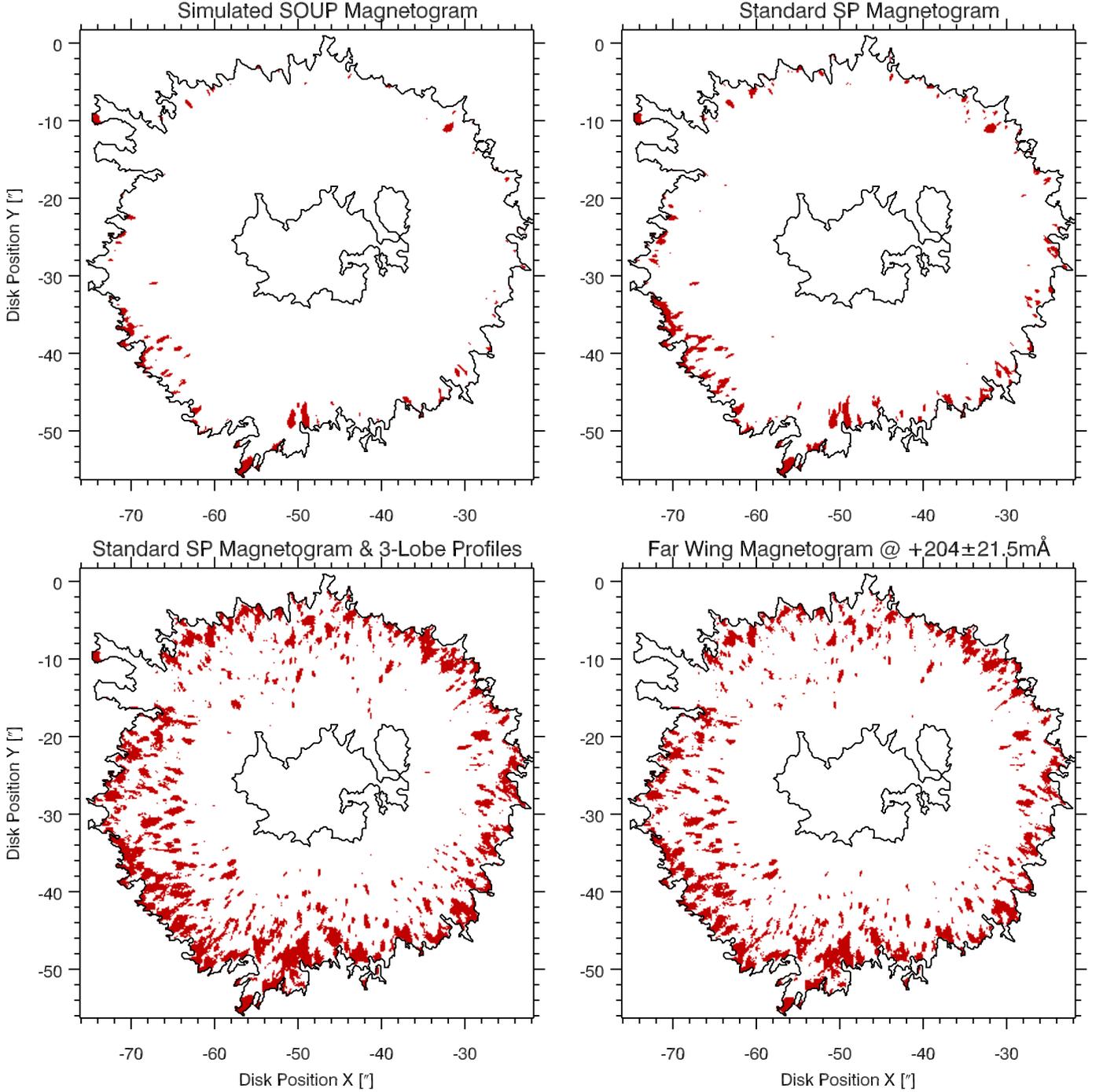}
\caption{Clockwise from top left: Penumbral magnetic fields of opposite polarity as they appear in a simulated SOUP@SST magnetogram, in a standard HINODE SP magnetogram, in a HINODE magnetogram computed in the far line wing of Fe I 630.25~nm, and in a standard HINODE magnetogram if 3-lobe Stokes V profiles are taken into account as well.}
\label{fig:Franz_fig05}
\end{center}
\end{figure*}

Magnetic fields of opposite polarity are commonly inferred from `normal' 2-lobe Stokes V profiles of opposite sign. Here, we demonstrated that opposite polarity in a particular pixel can also be inferred from the presence of 3-lobe profiles (if the spots is at disk center), leading to a much larger fraction of opposite polarities in a penumbra. This is of special relevance for penumbral models: \citet{2006A&A...447..343S}, for example, argue against `flux tube' models stating that these models need a large amount of submerging magnetic flux, which is not (or only partially) detected in magnetograms by \citet{2005A&A...436.1087L} \citep[cf. also][]{1993A&A...267..287S,1997Natur.389...47W}.
 
Here we show that there is much more return flux than detected with `Langhans'-magnetograms, if one takes into account 3-lobe profiles. In Sec.~\ref{subsec:langhans} we use our data to construct (i) magnetograms mimicking Langhans -- cf.~\ref{subsec:langhans}, (ii) compute magnetogram masks of V-profiles with opposite sign following Lites -- cf.~Sec.~\ref{subsec:lites}, (iii) and merge the latter mask with a mask of 3-lobe profiles to quantify all locations of opposite polarity -- cf.~Sect.~\ref{subsec:3-lobe}. Finally (Sec.~\ref{subsec:ichimoto}) we compare our mask to a map, in which only the V-signal in the far red wing is displayed.

\subsection{Simulated SOUP@SST magnetogram}
\label{subsec:langhans}

The study of \citet{2005A&A...436.1087L} was conducted at the Swedish Solar Tower (SST), using the Lockheed Solar Optical Universal Polarimeter (SOUP). This Lyot-type filtergraph measures the intensities of left and right circular polarized light with a bandpass of FWHM 7.2 pm at various wavelengths across the spectral line \citep{1981siwn.conf..326T}. In the case of Fe I 630.25, `Langhans'-magnetograms are constructed like
 
\begin{equation}
\rm{M}=\frac{\rm{I}_{\rm{blue}}^{\rm{LCP}}-\rm{I}_{\rm{blue}}^{\rm{RCP}}}{\rm{I}_{\rm{blue}}^{\rm{LCP}}+\rm{I}_{\rm{blue}}^{\rm{RCP}}}
\label{eq:magneto}
\end{equation}
 
\noindent with $\rm{I}_{\rm{blue}}^{\rm{LCP}}$ and $\rm{I}_{\rm{blue}}^{\rm{RCP}}$ being the intensities of left and right circular polarized light, measured by SOUP detuned $-5$ pm to the blue with respect to the linecore.
 
To simulate SST observations with the SOUP instrument, we first convolved the respective HINODE Stokes V measurements with a Gauss profile (100\% transmission at $-$5~pm from the line core of Fe I 630.25~nm with a FWHM of 7.2~pm). In a second step we integrated the obtained profiles over the full spectral range of SOUP and assigned a magnetic field of opposite polarity to the respective pixel if the result is the opposite sign to the umbral pixels.

Masking opposite polarity pixels with a red color, the morphology of this map is displayed in the upper left hand panel of Fig.~\ref{fig:Franz_fig05}. We see that opposite polarity pixels are predominantly detected close to the outer penumbral boundary. Only 2\% (1506 out of 74842 pixels) of the penumbral area is of opposite polarity. From these, 91\% are locations of downflows ($\rm{v}_{\rm{dop}} > 0.1$~km~s$^{-1}$). From all downflow pixels, 5\% (1350 out of 26064 pixels) have opposite polarity. 

\subsection{Standard SP magnetogram}
\label{subsec:lites}
 
Using the standard SP data pipeline (SSWIDL), maps of the apparent longitudinal magnetic field, $\rm{B}^{\rm{app}}_{\rm{lon}}$, can be obtained. According to \citet{1999ApJ...517.1013L,2008ApJ...672.1237L}\footnote{Even though the definition from $\rm{B}^{\rm{app}}_{\rm{lon}}$ in these publications is slightly different to Eq.~\ref{eq:bapp}, they yield identical maps of $\rm{B}^{\rm{app}}_{\rm{lon}}$ in the penumbra. Differences occur in the signal-to-noise ratio of weak Stokes V signals, which are not very common in the penumbra.} -- the apparent longitudinal magnetic field is proportional to
 
\begin{equation}
\rm{B}^{\rm{app}}_{\rm{lon}} \propto \frac{\int_{\uplambda_{\rm{b}}}^{\uplambda_0}\rm{V}(\uplambda)\rm{d}\uplambda-\int_{\uplambda_0}^{\uplambda_{\rm{r}}}\rm{V}(\uplambda)\rm{d}\uplambda}{\rm{I}_{\rm{c}}\int_{\uplambda_{\rm{b}}}^{\uplambda_r}\rm{d}\uplambda}
\label{eq:bapp}
\end{equation}
 
\noindent where $\rm{I_{c}}$ denotes the continuum intensity and ${\uplambda_0}$ the zero-crossing wavelength. The upper and lower limits of integration are represented by $\uplambda_{\rm{r}}$ and $\uplambda_{\rm{b}}$, which are set to $\uplambda_{\rm{r/b}}=\uplambda_0\pm$30~pm, respectively. The polarity of the flux in the maps of $\rm{B}^{\rm{app}}_{\rm{lon}}$ is thus determined from the sign of the area of the two lobes around the zero crossing. We manipulated the IDL routines in such a way that only the Fe I 630.25~nm line is considered. 

The corresponding `SP standard'-mask of opposite polarity is shown in the upper right hand panel of Fig.~\ref{fig:Franz_fig05}. As in the `Langhans'-mask, the pixels are located in the outer penumbra, but almost two times as many are found with this method: 4\% (2783 out 74842 pixels). From these, 92\% correspond to downflows. All opposite polarity pixels of the `Langhans'-mask are contained in this `SP standard'-mask. Now 10\% (2539 out of 26064 pixels) of all downflow pixels have opposite polarities.
 
\subsection{Standard SP magnetogram and 3-lobe profiles}
\label{subsec:3-lobe}

In this study we determine the amount of opposite polarities in the penumbra including 3-lobe Stokes V profiles. To this end we merge the `SP standard' mask with the mask of 3-lobe profiles (V$_{\rm{3-lobe}}$). This is justified since regular but mirrored Stokes V profiles remain undetected in the V$_{\rm{3-lobe}}$ mask (cf.~the criteria defined in Sec.~\ref{sec:obs}), but they are still tracers of opposite polarity.

Interestingly, the two sets are not disjunct. We find that only half of the profiles in the `SP standard' mask are mirrored Stokes V profiles. The other half contains 3-lobe profiles in which the additional lobe on the red side becomes larger than the difference in the areas of the two regular lobes. If Eq.~\ref{eq:bapp} is applied to the latter group of profiles, the result will be interpreted as magnetic fields of opposite polarity.

The superposition of V$_{\rm{3-lobe}}$ and the `SP standard' mask is shown in the bottom left hand panel of Fig.~\ref{fig:Franz_fig05}. Now 17\% (12594 out of 74842 pixels) of the penumbral area shows magnetic fields of opposite polarity and 87\% of these opposite polarities are located in downflows. Thus, the amount of all the penumbral downflows that harbor a magnetic field of opposite polarity increases to 40\% (10264 out of 26064 pixels).

In contrast to the two previous masks, the change in the morphology is striking -- cf. lower left hand panel in Fig.~\ref{fig:Franz_fig05}. The return flux is very prominent in the mid and outer penumbra, but occurs in the inner penumbra too. This is also the case in the penumbral side pointing towards disk center. This implies that the 3-lobe profiles we use as tracers of penumbral return flux cannot be due solely to horizontal Evershed motions.

\subsection{Magnetogram in the far line wing}
\label{subsec:ichimoto}

\citet{2007PASJ...59S.593I} computed magnetograms in the far red wing of the Fe 630.15 nm line to illustrate penumbral magnetic fields of opposite polarity. We followed this procedure, and computed a magnetogram at $+20.4$~pm from the line core of the Fe 630.25 nm line in order to compare such a measurement to our analysis. In contrast to \citet{2007PASJ...59S.593I}, we considered a Stokes V profiles as an indicator of opposite polarity only if the signal at $+20.4\pm2.15$~pm, i.e the neighboring spectral measurements, are above the 3$\upsigma$ noise level, too.

The lower right hand panel of Fig.~\ref{fig:Franz_fig05} shows a mask of opposite polarities in the far wing magnetogram. They occupy 13\% (9806 out of 74842 pixels) of the penumbral area and 85\% of them are located in downflows. As a result the fraction of penumbral downflows with magnetic fields of opposite polarity is 32\% (8364 out of 26064 pixels). Overall, the opposite polarity patches are smaller when compared to the previous mask, i.e. left hand panel of Fig.~\ref{fig:Franz_fig05}, but the overall morphology is very similar.

Our far wing magnetogram recovers 86\% of the opposite polarity pixels of the `SP standard'-magnetogram. Differences occur for example if the third lobe of a 3-lobe profile is not above the noise level at the specified wavelength. Furthermore, there are Stokes V profiles with only one or sometimes even with four lobes. We cannot rule out that such profiles yield a opposite polarity signal in the `SP standard' magnetogram, but at the same time, are not detected in the far wing magnetogram.

\begin{table}[h]
\begin{center}
	\caption{Penumbral return flux in different magnetograms.}
	\begin{tabular}{cccc}
		\hline
		\hline
		\\[-2ex]
		{Method of} & {Penumbral area with} &  {Fraction of return flux}\\
		{detection}&{opposite polarity[\%]} & {in downflows\footnotemark{} [\%]} \\
		\hline
		\\[-2ex]
		{Intensity at} & \multirow{2}{*}{2} & \multirow{2}{*}{5} \\
		{fixed wavelength} & & \\
		\\[-1.5ex]
		{Standard SP} & \multirow{2}{*}{4} & \multirow{2}{*}{10} \\
		{magnetogram}  & & \\
		\\[-1.5ex]
		{Far Wing} & \multirow{2}{*}{13} & \multirow{2}{*}{32} \\
		{magnetogram }  & & \\
		\\[-1.5ex]
		{SP Magnetogram} & \multirow{2}{*}{17} & \multirow{2}{*}{40}\\
		{and 3-lobe profiles}  & & \\
		\hline
	\end{tabular}
	\label{Tab_4}
\end{center}
\end{table}
\footnotetext{We consider only penumbral profiles with a downflow signal above the detection limit of 0.1~km~s$^{-1}$, cf.~\citet{2009A&A...508.1453F}. For Spot D these are 26064 out of 74842 penumbral profiles.}


\section{Discussion}
\label{sec:discuss}

The existence of penumbral downflows harboring magnetic fields of opposite polarity is an essential finding in the context of penumbral models. That these downflows, together with the opposite polarities, occur mainly in the middle and outer penumbra and that (at HINODE resolution) they appear patchy instead of elongated, supports the idea of flows in arching magnetic flux tubes.

Such flux tubes are a central ingredient of flux tube models \citep[cf.~e.g.][]{1968MitAG..25..194M,1997Natur.390..485M,2002AN....323..303S}. In the framework of the gappy penumbral model \citep{2006A&A...460..605S,2006A&A...447..343S}, the existence of opposite polarities as well their observed morphology cannot be explained.

\citet{2012ApJ...750...62R} argues that the majority of penumbral downflows advect magnetic field, which are still connected to the upper boundary of the simulation box, below the surface without changing its polarity. In any case, our findings prove that a large fraction of the emerging magnetic flux submerges within the penumbra.

\begin{figure}[h]
\begin{center}
    \includegraphics[width=\columnwidth]{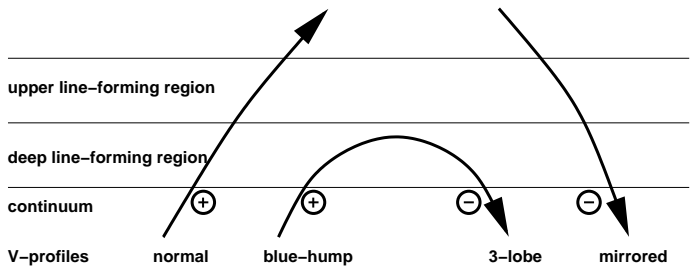}
\caption{Proposed configuration of the magnetic field in the line-forming layer of Fe I 630.25~nm.}
\label{fig:Franz_fig06}
\end{center}
\end{figure}

\paragraph{\bf Return flux:} We speculate that blue-hump and 3-lobe profiles indicate the presence of low-lying loops in the penumbra: the upflow footpoint is seen as a blue-hump profile, while the downflow footpoint (of opposite polarity) is seen as a 3-lobe profile. This is sketched in Fig.~\ref{fig:Franz_fig06}.
In this scenario, out of the 17\% of opposite polarity, 15\%  exist in the deep line-forming region only (3-lobe profiles), while 2\% fill the entire line-forming region (mirrored profiles). If we assume further that all the  submerging magnetic flux has a merging counterpart (blue-hump profile), 2$\,\cdot$15\% = 30\% of the penumbral pixels belong to magnetic field that returns in the deep photosphere. In this picture 2$\,\cdot$2\% = 4\% of the pixels show arches with apices in higher photospheric layers, and 66\% of the pixels harbor magnetic fields that continue beyond the line-forming region.

\paragraph{\bf 3-lobe profiles and upflows:} Interesting is that 3-lobe profiles sometimes exist in upflows (v$_{\rm{dop}} < - 0.1$~km s$^{-1}$) as well. These profiles look like a superposition of blue-hump and 3-lobe profiles. In our analysis these profiles occupy 1.3\% of the penumbral area, and they primarily occur between up- and downflows or they surround downflow patches. This suggests that such profiles are a mixture of up- and downflows that are not spatially resolved.

\paragraph{\bf Red-hump profiles:} \citet{2010A&A...524A..20K} found another type of asymmetric Stokes V profiles which show a hump on the red side of a regular V profile. These profiles would hint at magnetic fields of the same polarity in the downflow region. However, patches of red-hump profiles are rather small ($\approx 0.\arcsec5$) and far less abundant than the regions of 3-lobe profiles. Since these profiles occur co-spatially with a brightening in Ca II H, \citet{2010A&A...524A..20K} speculate that such red-hump profiles could be signs of magnetic reconnection.
 
\paragraph{\bf MIcro Structured Magnetic Atmospheres (MISMA):} \citet{2009A&A...508..963S} studied 3-lobe Stokes V profiles under the MISMA hypothesis \citep[cf. also][]{2005ApJ...622.1292S}. Similar to our results, these authors identify penumbral magnetic fields of opposite polarity using 3-lobe profiles. They find that a multiple component atmosphere is needed to properly model these asymmetric profiles. However, unlike our {\it two-layer atmosphere}, the MISMA hypothesis assumes that the solar atmosphere is structured on scales below the resolution limit of any observation available today \citep{1996ApJ...466..537S}. Even though our findings are not at odds with the MISMA  hypothesis, we favor an atmospheric model that is as simple as possible for explaining all (current) observational features, i.e. a two-layer atmosphere.
 
\paragraph{\bf Dark filaments and dark cores:} \citet{2009A&A...508..963S} suggest that the weakening of the magnetogram signal in dark cores \citep{2007ApJ...668L..91B,2008A&A...489..429V} is due to the mixture of small-scale magnetic fields of opposite polarity within penumbral filaments, cf.~\citet{2010mcia.conf..210S}. We do not find a significant correlation between opposite polarity (3-lobe Stokes V profiles) and dark-cored filaments using HINODE SP data of a penumbra at disk center. Only in spots off disk center, e.g. NOAA 10933 observed on January 7, 2007 at $\Uptheta\approx20^{\circ}$, such profiles show up in dark cores in the limb-side penumbra close to the apparent magnetic neutral line. Such behavior, however, is readily explained by the inclination of the flow channel (dark core) under the projection of the heliocentric angle. While the observed inclination (zenith plus heliocentric angle) of the magnetic field within the flow channel is larger than 90$^{\circ}$, this is not the case for the background atmosphere, which has a zenith angle smaller than 90$^{\circ}$. Thus, it is a projection effect that yields 3-lobe Stokes V profiles, but this does not indicate opposite polarities in the local reference frame. Because of the arguments made above, we follow the common interpretation that the weaker magnetogram signal of the penumbral `intra-spines' is due to the larger zenith angle of the magnetic field, causing a reduction in the Stokes V signal.
 
\paragraph{\bf Principal component analysis:} Our simple approach of counting the lobes of Stokes V profiles is limited by the signal-to-noise ratio of the HINODE SP. It was shown by \citet{2003A&A...412..875D} that all Stokes profiles can be interpreted as an infinite linear combination of a specific set of Hermite polynomials which form an orthonormal basis of squared integrable functions. Such a basis is noise free, and errors occur only because, in practice, an infinite number of terms cannot be considered in the series expansion. Using such a principal component analysis, it might be possible to detect an even larger number of magnetic fields with opposite polarity. 


\section{Summary and conclusion}
\label{sec:conclusions}

We explored strongly asymmetric Stokes V profiles in the penumbra of sunspots close to disk center, by using data of high spatial and spectral resolution from the Hinode spectropolarimeter. We can confirm earlier findings by \citet{2007PASJ...59S.593I} that penumbra up- and downflows are accompanied by strongly asymmetric Stokes V profiles. The profiles of upflows show an additional hump on the blue side, while the profiles from downflow regions contain an additional lobe on the red side, i.e., have three lobes.
 
Using the `synthesize module' of SIR, we modeled two simple configurations to emulate these asymmetries: Two components that are separated laterally (two-component) or vertically (two-layer). Both models are capable of qualitatively reproduceing the blue-hump and 3-lobe V profiles. For the 3-lobe profiles both models need magnetic fields of opposite polarity. However, only a vertical separation can produce an area asymmetry in V profiles. To quantitatively fit the profiles we performed an inversion with SIR. To minimize the degrees of freedom, we used a one-component model with gradients along the line-of-sight, which provides similar ingredients as the two-layer model to reproduce the area asymmetries.
We found that the proper reproduction of a 3-lobed profile, and its area asymmetry requires a strong magnetic field of opposite polarity and a downflow, both concentrated in the deep photosphere ($0 \le \rm{log}(\uptau) \le -1.5$).
 
Studying the shape of Stokes V profiles, we ordered them in bins of Doppler velocity, and calculated average profiles for each bin. From strong blue-shifts to strong red-shifts, the profiles transform smoothly from blue-hump profiles to 3-lobe profiles. This demonstrates that blue-humps correspond to upflows, while 3-lobe profiles correspond to s.
 
In `classical' magnetograms only 4\% of the pixels are of opposite polarity; i.e., these magnetograms miss a large fraction of opposite polarity. Merging the masks deduced from opposite sign of V and 3-lobed shapes, 17\% of all penumbral pixels are of opposite polarity. From these, 87\% correspond to downflows. 

Our numbers are only lower limits. With higher spatial resolution and a better signal-to-noise ratio, we expect that a larger amount of penumbral return flux to be detected.


\begin{acknowledgements}

We want to thank O. Steiner for fruitful discussions, J.M. Borrero for his help with the SIR code, and W. Schmidt for valuable comments on the manuscript. Part of this work was supported by the \emph{Deut\-sche For\-schungs\-ge\-mein\-schaft, DFG\/} project number Schl.~514/3-2. {\it{Hinode}} is a Japanese mission developed and launched by ISAS/JAXA, with NAOJ as domestic partner and NASA and STFC (UK) as international partners. It is operated by these agencies in cooperation with ESA and NSC (Norway).
\end{acknowledgements}

\bibliographystyle{aa}
\bibliography{franz2012}
\end{document}